\documentclass[a4paper]{jpconf}

\usepackage{graphicx}

\usepackage{amsmath,amssymb,amsbsy,amsfonts}

\usepackage[numbers,square]{natbib}

\begin{document}
\title{Extended models of gravity in SNIa cosmological data using genetic algorithms}

\author{O. L\'opez-Corona}

\address{Instituto de Astronom\'ia, Universidad Nacional Aut\'onoma de M\'exico, AP 70-264, Distrito Federal
04510, M\'exico}

\ead{olopez@astro.unam.mx}

\begin{abstract}
In this talk I explained briefly the advantages of using genetic algorithms on any measured data but specially astronomical ones. This kind of algorithms are not only a better computational paradigm, but they also allow for a more profound data treatment enhancing theoretical developments. As an example, I will use the SNIa cosmological data to fit the extended metric theories of gravity of Carranza et al. (2013, 2014) showing that the best parameters combination deviate from theoretical predicted ones by a minimal amount. This means that these kind of gravitational extensions are statistically robust and show that no dark matter and/or energy is required to explain the observations.
\end{abstract}

\section{Introduction}
A Genetic Algorithm (GA) is an evolutionary based stochastic search algorithm that 
in some sense mimic natural evolution. Points in the search space are 
considered as individuals (solution candidates), which form a population.
The particular fitness of an individual is a number, indicating their quality for the problem at hand. As in nature, GAs include a set of fundamental genetic operations that work on the genotype (solution candidate codification). A concrete GA is specified by a particular mutation, recombination, or selection operators \cite{Mitchell-98}.

These algorithms operate with a population of individuals $P(t) = {x_{1}^{t}, ...,x_{N}^{t}}$, 
for $t$ iteration, where the fitness of each $x_{i}$ individual is evaluated
according to a set of objective functions $f_{j}(x_{i})$. This objectives functions allows to 
order from best to worst individuals of the population in a continuum of degrees of adaptation.
Then, individuals with higher fitness, recombine their genotypes to form the gene pool of the next 
generation, in which random mutations are also introduced to produce new variability.

 A fundamental advantage of GA versus traditional methods is that GA’s solve discrete,
nonconvex, discontinuous, and non-smooth problems successfully \cite{Lopez-Corona-13},
and thus they have been widely used in Ecology, Natural Resources Management, among other fields, 
but not so much in astrophysics. Nevertheless they have been recently used by Nesseris (2011)
for parameter search in $\Lambda CDM$ models with SNIa data \cite{Nesseris}.

When Zwicky applied the virial theorem to the cluster of galaxies COMA, he obtained some results 
suggesting the existence of additional  mass 
that was not visible. Based on the motions of galaxies 
near its edge, Zwicky estimated the total mass of the cluster should be about 400 times more than expected. 
The gravity of the visible galaxies in the 
cluster was very little to account for the actually observed galaxy speed, 
so it should be much more. This is known as the "missing mass problem". 

In the 1960s and 1970s, there were measurements of the rotation curves of spiral galaxies with a 
higher degree of accuracy than anything previously achieved before, that supported Zwicky findings. These results showed that many stars in different orbits of spiral galaxies rotate at about the same angular velocity, 
implying that their densities were very uniform beyond the location of many of the stars (the galactic bulge). 
This result suggests that even Newtonian gravity does not apply universally or that, conservatively, over 
50\% of the mass of galaxies was contained in the relatively dark galactic halo. This 
non-baryonic 
interacting matter was termed as Dark Matter \cite{Sanders-10,Bachall-04,Trimble-87}. 

A few years later it was discovered that the Universe is undergoing an accelerated expansion \cite{Riess-04,Spergel-07,ReadHead}.
This acceleration is usually attributed either to a cosmic fluid with negative pressure dubbed Dark Energy 
or to an IR modification of gravity. As pointed out by Mendoza (2014), it seems that ideas of introducing occult 
entities to describe physical phenomena reoccurs from time to time.

Modified theories of gravity have received increased attention lately due to combined motivation
coming from high-energy physics, cosmology and astrophysics. Among numerous alternatives
to Einstein’s theory of gravity, theories which include higher order curvature invariants, and
specifically the particular class of $F(R)$ theories, have a long history \cite{Sotiriou-10}. 
(see e.g. Mariana Espinosa talk about the Gravity Apple Tree and the many theories of gravity developed after General Relativity)










Bernal et. al \cite{Bernal11} 
and Mendoza \cite{Mendoza13} 
have shown that a useful way to write the gravitational actions $S_{g}$ is given by:

\begin{equation}
S_{g}=-\frac{c^{3}}{16\pi G}\int d^{4}x \sqrt{-g} \frac{f(\chi)}{L^{2}}
\end{equation}

\noindent where the dimensionless Ricci scalar ir given by $\chi:=L^{2}R$, for the Ricci scalar $R$. In the previous equation, $c$ is the speed of light, $G$ Newton's gravitational constant and $g$ is the determinant of the space-time metricc. Note that the function $f(\chi)=\chi$ recovers general relativity and for the case of $f(\chi)\not=\chi$ the "coupling factor" with dimensions of length $L$ makes that equation dimensionally correct.

As explained in Mendoza (2014) there are two cases that can yield a MONDian behaviour:

\begin{equation}
f(\chi)= \begin{cases} \chi^{\frac{3}{2}} 
\\ 
\chi^{-3} 
\end{cases} 
\end{equation}
 
\noindent both reproduce flat rotational curves and the Tully-Fisher relation and show the fact that the energy-momentum tensor - or the matter Langrangian - is somewhat 
embedded into the gravitational action. We only 
on the description based on $f(R,T)$ by Harko et al.
and covered by Carranza et al. \cite{Carranza13}, but also see Mendoza \cite{Mendoza13} and the proceedings by S. Mendoza in this collection.

It has been shown that the interpretation of the SNIa data depends greatly on the type of parametrization
used to perform a data fit.  For this reason, choosing a priori a model can thus adversely
affect the validity of the fitting method and lead to compromised or misleading results. Then a very 
important complication in the investigation of the behavior of dark energy or for that matters,
its counterpart of extended models of gravity, occurs due to the bias introduced by the parameterizations used \cite{Nesseris}.










\section{Methods}

We conceptualized the parameter calibration as an optimization problem and propose to resolve it
using both Monte Carlo (MC) techniques and Genetic Algorithms. 

As it is well known from Taylor series, any (normal) function may be approximated using a polynomial of the correct order. Of course, even when this is mathematical correct no one would consider that a polynomial is a universal model for every physical phenomenon. In that line of thought, one may fit any data using a model with many free parameters, and even when this model may have a great performance  in 
a statistical sense, it could be at the same time an incorrect model in the physical perspective.

In these sense, an important question is, How much better must a complex (more parametrers) model fit before we say that the extra parameter is necessary? or in a more straight forward, How do we trade off fit with simplicity? That has been the motivation in recent year for new model selection criteria developement in statistics, all of which define simplicity in terms of the paucity of parameters, or the dimension of a model (see Forster and Sober, 1994, for a non-technical introduction). These criteria include Akaike’s Information Criterion (AIC) \cite{Akaike-74,Akaike-85}, the Bayesian Information Criterion (BIC) \cite{Schwarz-78}, and Minimum Description Length (MDL) \cite{Rissanen-89}. They trade off simplicity and fit a little differently, but all of them address the same problem as significance testing: Which of the estimated "curves" from competing models best represent reality? This work has led to a clear understanding of why this form of simplicity is relevant that question \cite{Forster-94}.

Akaike (Akaike-74,Akaike-85) has shown that choosing the model with the lowest expected information loss (i.e., the model that minimizes the expected Kullback– Leibler discrepancy) is asymptotically equivalent to choosing a model $M_{j}$, from a set of models $i=1,2,...,k$  that has the lowest AIC value:

\begin{equation}
AIC=-2Log(\mathcal{L}_{j})+2V_{j},
\end{equation}

\noindent where $\mathcal{L}_{j}$, is the maximum likelihood for the candidate model and is determined by adjusting the $V_{j}$ free parameters in such a way as to maximize the probability that the candidate model has generated the observed data. This equation shows that AIC rewards descriptive accuracy via the maximum likelihood , and penalizes lack of parsimony according to the number of free parameters (note that models with smaller AIC values are to be preferred). In that sense, Akaike extended this paradigm by considering a framework in which the model dimension is also unknown, and must therefore be determined from the data. Thus, Akaike proposed a framework wherein both model estimation and selection could be simultaneously accomplished. For those reasons, AIC is generally regarded as the first, most widely known and used model selection tool. 

Taking as objective function the AIC, we performed a Monte Carlo optimization process using an R script. For the GA analysis we used a modified versión of Kumara Sastry code in C++ \cite{Kumara}. For both Monte Carlo and GA, we take averages over 1000 independent sets of parameters each one with 1000 iterations or generation respectively.

\section{Results}

The results are quite straightforward and show consistently that GA are much more better that the other parameter search techniques used. For example, GA for $\chi^{3/2}$ is of the order of forty percent better than Monte Carlo. In Fig.1 we see data from SNIa, predicted values of the model using theoretical set of parameters in green, values obtained using MC parameter search in blue, and values obtained using GA as search technique in red. This performance was accomplished by desegregating the original data into three subsets and defining an independent AIC objective function for each one, so we make sure that differences in data density over z values does’t interfere with parameter estimation.

\begin{figure}
\begin{center}
\includegraphics[scale=0.8]{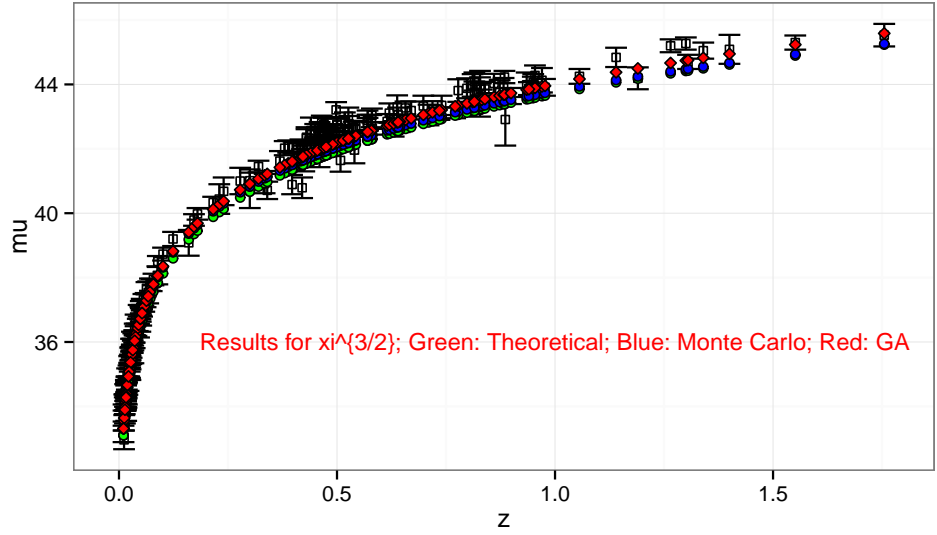}
\caption{Comparative of optimization techniques for $\chi^{3/2}$ model. Predicted values of the model using theoretical set of parameters in green, values obtained using MC parameter search in blue, and values obtained using GA as search technique in red. }
\par\end{center}
\end{figure}

As shown in Fig.2, $\chi^{-3}$ we have the same result, with GA as the best parameter optimization option. It is important to point out that the estimation $b=-3.0001\pm0.0015$, implies that the Hubble constant $h=0.6919\pm0.0035$ by the GA is consistent with standard values used in the concordance $\Lambda CDM$ model.

\begin{figure}
\begin{center}
\includegraphics[scale=0.8]{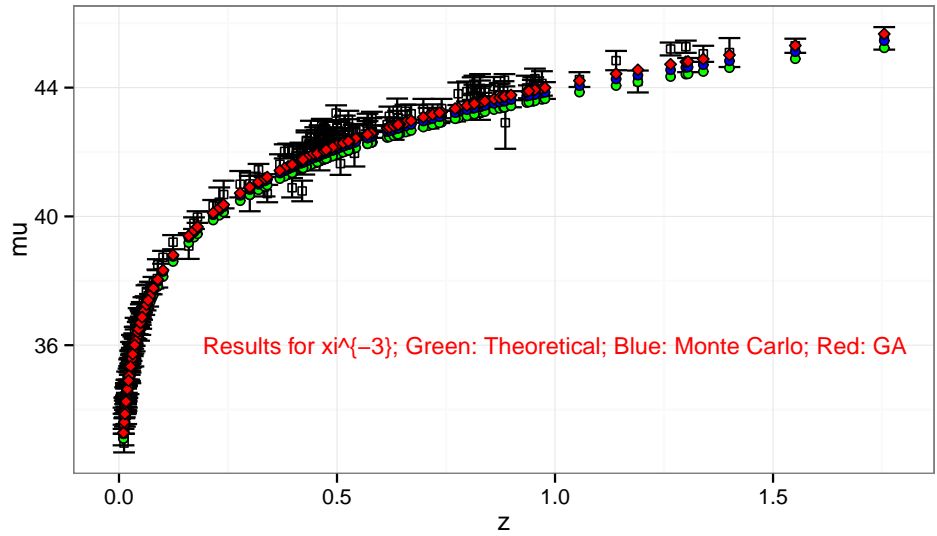}
\caption{Comparative of optimization techniques for $\chi^{-3}$ model. Predicted values of the model using theoretical set of parameters in green, values obtained using MC parameter search in blue, and values obtained using GA as search technique in red. }
\par\end{center}
\end{figure}

Comparing $\chi^{-3}$ and $\chi^{3/2}$, it turns out that $\chi^{-3}$ result better but just for a 1.3\% (Fig.3). In this range of z, $\chi^{3/2}$  and $\chi^{-3}$ are almost indistinguishable but $\chi^{3/2}$ is base in a punctual concept mass (dust type of models) meanwhile $\chi^{-3}$  use a density concept, allowing to apply it to more astrophysical systems than the former.

\begin{figure}
\begin{center}
\includegraphics[scale=0.8]{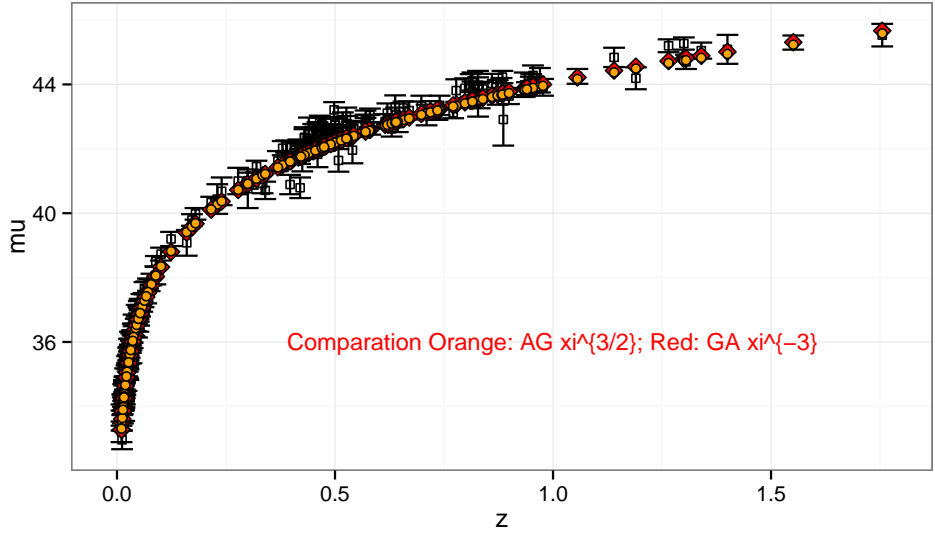}
\caption{Comparative results
of $\chi^{3/2}$ and  $\chi^{-3}$ model. In this range of redshift 
z, $\chi^{3/2}$  and $\chi^{-3}$ are almost indistinguishable}
\par\end{center}
\end{figure}

\section{Discussion and Conclusions}

Genetic Algorithms (GA) have proven to be an efficient technique for parameter estimation
in cosmological data, but as proposed models performance are almost the same with the optimized set of parameters. Which is then the best model?
In order to respond this critical question we need to work with data in a far as $z\ge 10$ 
(perhaps using gamma ray
data) where the differences between models are expected to be higher.

Another path to overcome this statistical indistinguishability is to incorporate to the current data driven criteria a physics driven one, for model selection. This new critiria could be based on information-thermodynamical principles; and use a muti-data optimization process (which include different z regimes and observed objects); or a
mix of multi-data and multi-criteria optimization process, that would allow us to really exploit the real power 
of GA by incorporating a multi-objective optimization approach. The point is to have a set of various databases 
representing different aspects or regimes of a physical phenomena and then estimate the parameters of a model minimizing at the same time

Finally, with this results in mind we are now in position to compare the $\chi^{-3}$ model with the standard 
$\Lambda CDM$ model.

\section*{Acknowledgments}

This work was supported by a DGAPA-UNAM posdoctoral fellowship at Instituto de Astronom\'ia and
the grant IN111513-3; it was also supported by CONACyT-SNI 62929 and grant CB-2009-01, 132400)

\bibliographystyle{iopart-num}

\end{document}